\documentstyle[12pt,a4wide,overcite,epsfig]{article}

\newcommand{\Preprint}{\vspace*{-1.0cm} \noindent hep-ph/9610243
\hfill 
  FTUV/96-65 \\ \mbox{}\hfill IFIC/96-74 \\  \mbox{}\hfill
  October 1996 \\ }       
\def\Thebibliography#1{\section*{References}\list
{\arabic{enumi}.}{\settowidth\labelwidth{#1.}\leftmargin\labelwidth
 \advance\leftmargin\labelsep
 \usecounter{enumi}}
 \def\newblock{\hskip .11em plus .33em minus .07em}
 \sloppy\clubpenalty4000\widowpenalty4000
 \sfcode`\.=1000\relax}

\def\refjl#1#2#3#4#5#6{\bibitem{#1} #2, {\it #3 \/} {\bf #4} (#5) #6.}
\def\refbk#1#2#3#4{\bibitem{#1} #2, {\it #3}, #4.}
\def\etal{{\it et al\/}}
\def\NP{Nucl. Phys.}
\def\PL{Phys. Lett.}
\def\PRL{Phys. Rev. Lett.}
\def\PR{Phys. Rev.}

\def\ZP{Z. Phys.}

\def\NC{Nuovo Cimento}
\def\RMP{Rev. Mod. Phys.}
\def\RPP{Rep. Prog. Phys.}

\def\PTP{Progr. Theor. Phys.}
\def\PPNP{Prog. Part. Nucl. Phys.}
\newcommand{\newsec}[1]{\section{#1}}
\newcommand{\eqn}[1]{(\ref{#1})}
\newcommand{\bel}[1]{\be\label{#1}}
\newcommand{\be}{\begin{equation}}
\newcommand{\ee}{\end{equation}}
\newcommand{\ba}{\begin{array}{c}}
\newcommand{\bat}{\begin{array}{cc}}
\newcommand{\ea}{\end{array}}
\newcommand{\beqn}{\begin{eqnarray}}
\newcommand{\eeqn}{\end{eqnarray}}

\newcommand{\bi}{\begin{itemize}}
\newcommand{\ei}{\end{itemize}}

\newcommand{\cL}{{\cal L}}
\newcommand{\cH}{{\cal H}}

\newcommand{\cM}{{\cal M}}
\newcommand{\cO}{{\cal O}}
\newcommand{\cP}{{\cal P}}

\newcommand{\no}{\nonumber}
\newcommand{\lsim}{\stackrel{<}{_\sim}}
\newcommand{\rms}{\rm\scriptsize}

\newcommand{\Or}{$O$}
\begin{document}
\thispagestyle{empty}
\begin{titlepage} 
\Preprint
\mbox{}
\vspace{7cm}
\begin{center}
{\large\bf RARE KAON DECAYS\footnote{Invited Talk at the
  Workshop on K Physics, Orsay, France, May 30 -- June 4, 1996}}
   \\[2\baselineskip]
  {{\bf A. Pich} \\ 
  {\it Departament de F\'{\i}sica Te\`orica, IFIC,  
  Universitat de Val\`encia -- CSIC \\ 
  E-46100  Burjassot, Val\`encia, Spain}} 
\vspace{5cm}\\ 
{\bf Abstract} 
\end{center}
\noindent
Rare K decays are an important testing ground of the
electroweak flavour theory. They can provide new signals
of CP--violation phenomena and, perhaps, a window into physics
beyond the Standard Model.
The interplay of long--distance QCD effects in strangeness--changing
transitions can be analyzed with Chiral Perturbation Theory techniques. 
Some theoretical predictions obtained within this
framework for radiative kaon decays are reviewed, together with the
present experimental status. 

\vfill
\end{titlepage}
\newsec{Introduction}

High--precision experiments on rare kaon decays offer the exciting
possibility of unravelling new physics beyond the Standard Model.
Searching for forbidden flavour--changing processes
\cite{LI:96,WI:96}     
at the $10^{-10}$ level
[Br$(K_L\to\mu e) < 3.3\times 10^{-11}$,
Br$(K_L\to\pi^0\mu e) < 3.2\times 10^{-9}$,
Br$(K^+\to\pi^+\mu e) < 2.1\times 10^{-10}$, \ldots],
one is actually exploring energy scales above the 10 TeV
region. The study of allowed (but highly suppressed) decay modes
provides, at the same time, very interesting tests of the Standard
Model itself. Electromagnetic--induced non-leptonic weak transitions
and higher--order weak processes are a useful tool to improve our
understanding of the interplay among electromagnetic, weak and strong
interactions. In addition, new signals of CP violation, which would
help to elucidate the source of CP--violating phenomena, can be looked
for.
 
    Since the kaon mass is a very low energy scale, the theoretical
analysis of non-leptonic kaon decays is highly non-trivial. While the 
underlying flavour--changing weak transitions among the constituent 
quarks are associated with the $W$--mass scale, the 
corresponding hadronic amplitudes are governed by the long--distance
behaviour of the strong interactions, 
i.e. the confinement regime of QCD.

The standard short--distance approach to weak transitions makes
use of the asymptotic freedom property of QCD
to successively integrate out the fields with heavy masses down to 
scales $\mu < m_c$.
Using the operator product expansion (OPE) and
renormalization--group techniques, one gets an effective $\Delta S=1$
hamiltonian,
\bel{eq:sd_hamiltonian}
\cH_{\mbox{\rms eff}}^{\Delta S=1} \, = \, {G_F\over\sqrt{2}}
V_{ud}^{\phantom{*}} V_{us}^*\,
\sum_i C_i(\mu) Q_i \, + \, \mbox{\rm h.c.},
\ee
which is a sum of local four--fermion operators $Q_i$,
constructed with the light degrees of freedom
($u,d,s; e,\mu,\nu_l$), modulated by
Wilson coefficients $C_i(\mu)$
which are functions of the heavy ($W,t,b,c,\tau$) masses.
The overall renormalization scale $\mu$
separates the short-- ($M>\mu$) and long-- ($m<\mu$) distance
contributions,  which are contained in $C_i(\mu)$
and $Q_i$, respectively.
The physical amplitudes are of course independent of $\mu$;
thus, the explicit scale (and scheme) dependence of the Wilson
coefficients, should cancel exactly with the corresponding dependence
of the $Q_i$ matrix elements between on--shell states.

Our knowledge of the $\Delta S=1$ effective hamiltonian has improved
considerably in recent years, thanks to the completion of the
next-to-leading logarithmic order calculation of the Wilson
coefficients \cite{buras}.
All gluonic corrections of $\cO(\alpha_s^n t^n)$ and 
$\cO(\alpha_s^{n+1} t^n)$ are already known,
where $t\equiv\log{(M/m)}$ refers to the logarithm of any ratio of 
heavy--mass scales ($M,m\geq\mu$). Moreover, the full $m_t/M_W$ dependence
(at lowest order in $\alpha_s$) has been taken into account.

Unfortunately, in order to predict the
physical amplitudes one is still confronted with the calculation of
the hadronic matrix elements of the quark operators.
This is a very difficult problem, which so far remains unsolved. 
We have only been able to obtain rough estimates using
different approximations (vacuum saturation, $N_C\to\infty$ limit, 
QCD low--energy effective action, \ldots)
or applying QCD techniques (lattice, QCD sum rules) which suffer from
their own technical limitations.

\begin{figure}[tbh]      
\setlength{\unitlength}{0.8mm} \centering
\begin{picture}(165,120)
\put(0,0){\makebox(165,120){}}
\thicklines
\put(10,105){\makebox(40,15){\large\bf Energy Scale}}
\put(58,105){\makebox(36,15){\large\bf Fields}}
\put(110,105){\makebox(40,15){\large\bf Effective Theory}}
\put(8,108){\line(1,0){149}} {\large
\put(10,75){\makebox(40,27){$M_W$}}
\put(58,75){\framebox(36,27){$\ba W, Z, \gamma, g \\
     \tau, \mu, e, \nu_i \\ t, b, c, s, d, u \ea $}}
\put(110,75){\makebox(40,27){Standard Model}}

\put(10,40){\makebox(40,18){$\lsim m_c$}}
\put(58,40){\framebox(36,18){$\ba  \gamma, g  \; ;\; \mu ,  e, \nu_i  
             \\ s, d, u \ea $}} 
\put(110,40){\makebox(40,18){$\cL_{QCD}^{N_f=3}$, \  
             $\cH_{\mbox{\rms eff}}^{\Delta S=1,2}$}}

\put(10,5){\makebox(40,18){$M_K$}}
\put(58,5){\framebox(36,18){$\ba\gamma \; ;\; \mu , e, \nu_i  \\ 
            \pi, K,\eta  \ea $}} 
\put(110,5){\makebox(40,18){ChPT}}
\linethickness{0.3mm}
\put(76,37){\vector(0,-1){11}}
\put(76,72){\vector(0,-1){11}}
\put(80,64.5){OPE} 
\put(85,27.5){\Huge ?} }
\end{picture}
\caption{Evolution from $M_W$ to the kaon mass scale.
  \label{fig:eff_th}}
\end{figure}
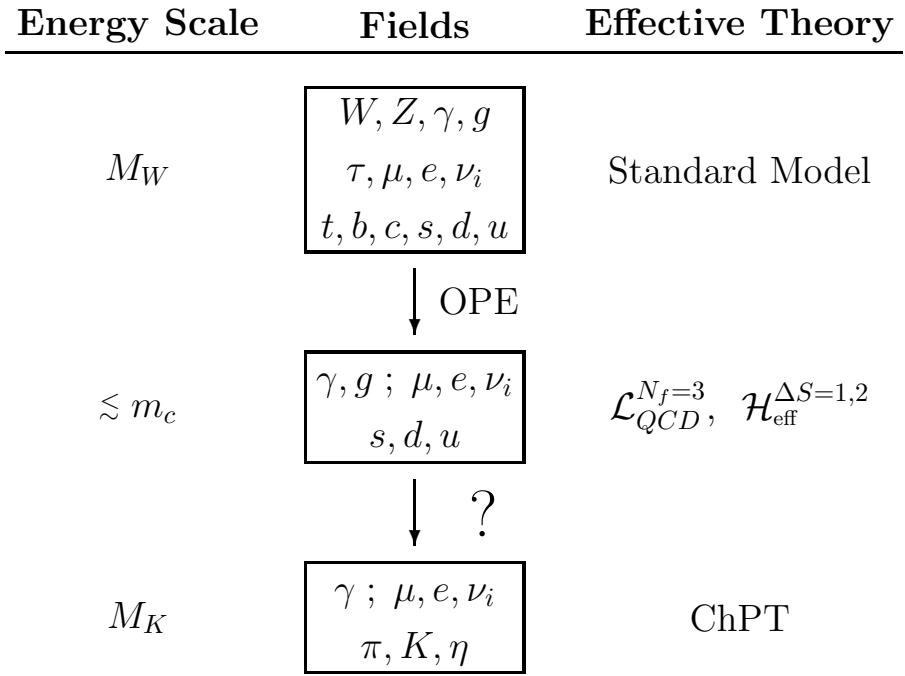

   Below the resonance region ($\mu < M_\rho$) the strong interaction
dynamics can be better understood with global symmetry considerations. 
We can take  advantage of the fact that the
pseudoscalar mesons are the lowest energy modes of the hadronic
spectrum: they correspond to the octet of Goldstone bosons associated
with the dynamical chiral symmetry breaking of QCD, $SU(3)_L\otimes
SU(3)_R \rightarrow SU(3)_V$.
The low--energy implications of the QCD symmetries can then be worked out
through an effective lagrangian containing only the Goldstone modes.
The effective chiral perturbation
theory  \cite{WE:79,GL:85,EC:95,PI:95} (ChPT) 
formulation of the Standard Model is an ideal
framework to describe kaon decays \cite{dR:95,EPR:96}. 
This is because in K decays the
only physical states which appear are pseudoscalar mesons, photons and
leptons, and because the characteristic momenta involved are small
compared to the natural scale of chiral symmetry breaking 
($\Lambda_\chi\sim 1$~GeV).

Fig.~\ref{fig:eff_th} shows a schematic view of the procedure used
to evolve down from $M_W$ to the kaon mass scale.
At the different energy regimes one uses different effective theories,
involving only those fields which are relevant at that scale.
The corresponding effective parameters (Wilson coefficients, chiral
couplings) encode the information on the heavy degrees of freedom 
which have been integrated out.
These effective theories are convenient realizations of the fundamental
Standard Model at a given energy scale 
(all of them give rise to the same generating functional and therefore to
identical predictions for physical quantities). From a technical point of
view, we know how to compute the effective hamiltonian at the charm--mass
scale. Much more difficult seems the attempt to derive the chiral
lagrangian from first principles. The symmetry considerations only fix
the allowed chiral structures, at a given order in momenta, but leave
their corresponding coefficients completely undetermined.
The calculation of the chiral couplings from
the effective short--distance hamiltonian, remains the main open problem
in kaon physics.

%
%
\newsec{Chiral Perturbation Theory}
 
   In the absence of quark masses, the QCD lagrangian is invariant under
independent $SU(N_f)$  
transformations of the left-- and
right--handed quarks in flavour space [$q_L \rightarrow g_L q_L$,
$q_R \rightarrow g_R q_R$, $g_{L,R}\in SU(N_f)_{L,R}$]. 
This $SU(N_f)_L\otimes SU(N_f)_R$ chiral symmetry, which should be
approximately good in the light quark sector (u,d,s), 
is however not seen in the hadronic spectrum: although hadrons can be 
nicely classified in $SU(3)_V$ representations,
degenerate multiplets with opposite parity do not exist.
To be consistent with this experimental fact, the ground state of the
theory (the vacuum) should not be symmetric under the chiral group.
The $SU(3)_L \otimes SU(3)_R$ symmetry spontaneously breaks down to
$SU(3)_{L+R}$ and, according to Goldstone's theorem, 
an octet of pseudoscalar massless bosons appears in the theory. 
The eight lightest hadronic states
($\pi^+,\pi^-,\pi^0,\eta,K^+,K^-,K^0$ and $\bar{K}^0$)
are then identified with the Goldstone bosons of chiral symmetry;
their small masses being generated by the quark mass matrix, which 
explicitly breaks the global symmetry of the QCD lagrangian.
 
The Goldstone nature of the pseudoscalar mesons implies strong constraints
on their interactions, which can be most easily analyzed on the basis of 
an effective lagrangian. 
Since there is a mass gap separating the pseudoscalar octet from the rest
of the hadronic spectrum, we can build an effective field theory containing
only the Goldstone modes.
The quark and gluon fields of QCD are replaced by a unitary
$3\times 3$ matrix \ $U(\phi) \equiv \exp(i \sqrt{2} \Phi /f)$,
incorporating the pseudoscalar octet fields:
\bel{eq:phi}
\Phi (x) \equiv {\vec{\lambda}\over\sqrt 2} \, \vec{\phi}
\, = \, 
\pmatrix{{1\over\sqrt 2}\pi^0 \, + 
\, {1\over\sqrt 6}\eta_8
 & \pi^+ & K^+ \cr
\pi^- & - {1\over\sqrt 2}\pi^0 \, + \, {1\over\sqrt 6}\eta_8   
 & K^0 \cr K^- & \bar K^0 & - {2 \over\sqrt 6}\eta_8 }.
\ee
$U^{ij}(\phi)$ parametrizes the Goldstone excitacions over the vacuum quark
condensate  $\langle\bar{q}_L^j q_R^i\rangle$. 
Under the chiral group, it transforms
as \ $U \rightarrow g_R U g_L^+$.   
 
To get a low--energy effective lagrangian realization of QCD, for the
light--quark sector, we should write 
the most general lagrangian involving the matrix
$U(\phi)$, which is consistent with chiral symmetry.
The lagrangian can be organized in terms of increasing powers of
momentum or, equivalently, in terms of increasing number of derivatives:
\bel{eq:leff}
\cL_{\mbox{\rms eff}}(U) = \sum_n \cL_{2n} \, .
\ee
In the low--energy domain we are interested in, the terms with a minimum
number of derivatives will dominate.
 
The lowest--dimensional effective chiral lagrangian is uniquely given by
\bel{eq:l2}
\cL_2 = {f^2\over 4} \left(\langle D_\mu U D^\mu U^\dagger\rangle + 
2 B_0 \,\langle\cM U^\dagger + U \cM\rangle \right),
\ee
where $f \simeq f_{\pi}=92.4$ MeV  is the pion decay constant (to lowest 
order), $\langle\ \rangle$ denotes the trace of the corresponding matrix 
and the covariant derivative
\bel{eq:cov}
D_{\mu}U=\partial _{\mu}U - i r_\mu U + i U l_\mu ,
\ee
accounts for the coupling to electromagnetism 
(and the weak gauge bosons),
\bel{eq:l_r}
r_\mu \equiv v_\mu + a_\mu = e Q A_\mu + \cdots \qquad\qquad
l_\mu \equiv v_\mu - a_\mu = e Q A_\mu + \cdots 
\ee
with the charge matrix
$Q={1\over 3} \,\mbox{\rm diag}(2,-1,-1)$.

The second term in \eqn{eq:l2} is an explicit breaking of chiral symmetry 
due to the presence of the quark mass matrix 
$\cM=\mbox{\rm diag}(m_u,m_d,m_s)$
in the QCD lagrangian.
The parameter $B_0$  ($\simeq -<\bar{u} u>/f^2$)
relates the squares of the
pseudoscalar meson masses to the quark masses,
\bel{eq:GMO}
B_0={M_{\pi^+}^{\ 2}\over m_u+m_d}
 = {M_{K^+}^{\ 2} \over m_u+m_s}
= {M_{K^0}^{\ 2} \over m_d + m_s}\, .
\ee
 
The effect of strangeness--changing non-leptonic
weak interactions with $\Delta S=1$ is incorporated as a perturbation to
the strong effective lagrangian $\cL_{\mbox{\rms eff}}$. 
At lowest order in the number of derivatives,
the most general effective bosonic lagrangian, with the same
$SU(3)_L\otimes SU(3)_R$ transformation properties as the short--distance
hamiltonian \eqn{eq:sd_hamiltonian}, contains two terms:
\bel{eq:lg8_g27}
\cL_2^{\Delta S=1} = -{G_F \over \sqrt{2}}  V_{ud}^{\phantom{*}} V_{us}^*
\left\{ g_8  \,\langle\lambda L_{\mu} L^{\mu}\rangle \  +
g_{27} \left( L_{\mu 23} L^\mu_{11} + {2\over 3} L_{\mu 21} L^\mu_{13}
\right) +
\mbox{\rm h.c.} \right\} ,
\ee
where the matrix $L_{\mu}=i f^2 U^\dagger D_\mu U$  represents the octet of 
$V-A$
currents, and $\lambda\equiv (\lambda_6 - i \lambda_7)/2$ projects onto the
$s\to d$ transition [$\lambda_{ij} = \delta_{i3}\delta_{j2}$].
The chiral couplings $g_8$ and $g_{27}$ measure the strength of the two
parts of the effective hamiltonian \eqn{eq:sd_hamiltonian} transforming as 
$(8_L,1_R)$ and $(27_L,1_R)$, respectively, under chiral rotations.
Their values can be extracted from $K \rightarrow 2 \pi$ decays
\cite{PGR:86}:
\bel{eq:g8_g27}
\vert g_8 \vert \simeq 5.1\, , \qquad\qquad 
\vert g_{27}/ g_8 \vert \simeq 1/18 \, .
\ee
The huge difference between these two couplings\footnote{
The $O(p^4)$ corrections \protect\cite{KMW:91,KDHMW:92}
give a sizeable constructive contribution to
the octet decay amplitude, which results in a 30\% smaller fitted value 
for $|g_8|$. 
}
shows the well--known
enhancement of the octet $\vert\Delta I\vert = 1/2$ transitions.
 
Using the lagrangians \eqn{eq:l2}  and \eqn{eq:lg8_g27}, 
the rates for decays 
like $K \rightarrow 3 \pi$ or $ K \rightarrow \pi \pi \gamma $ can be 
predicted
at $O(p^2)$ through a trivial tree--level calculation. However, the data
are already accurate enough for the next--order corrections to be sizeable.
Moreover, due to a mismatch between the minimum number of powers of momenta
required by gauge invariance and the powers of momenta that the 
lowest--order
effective lagrangian can provide \cite{EPR:87a,EPR:87b,EPR:88}, 
the amplitude for any non-leptonic 
radiative
K decay with at most one pion in the final state ($K \rightarrow \gamma
\gamma  , K \rightarrow \gamma l^+ l^- , K \rightarrow \pi \gamma \gamma ,
K \rightarrow \pi l^+ l^-$, \ldots)
vanishes to $O(p^2)$.
These decays are then sensitive to the non-trivial quantum field theory
aspects of ChPT.
 
At the one--loop level, corresponding to $O(p^4)$, we need to add to the
effective lagrangian all possible terms with four powers of momenta,
satisfying the symmetry constraints. Each term will introduce an additional
coupling constant, not fixed by chiral symmetry. These constants can be
seen as remnants of the fundamental theory after quarks and gluons have
been integrated out; they contain both long-- and short--distance 
information,
and some of them (like $g_8$) have in addition a CP--violating imaginary
part. Since the one--loop divergences are reabsorbed by the $O(p^4)$
couplings, these constants will depend, in general, on an arbitrary
renormalization scale.
 
The complete list of $O(p^4)$ terms describing strong and electromagnetic
interactions can be found in ref.~\citen{GL:85},
where the numerical values of the corresponding couplings have been
determined using experimental information.
Two of those terms are relevant for our purposes: 
%
\bel{eq:l4}
\cL_4^{\mbox{\rms em}} =  - i e L_9 F^{\mu\nu}
\,\langle Q D_\mu U D_\nu U^\dagger + Q D_\mu U^\dagger D_\nu U\rangle
+ e^2 L_{10} F^{\mu\nu} F_{\mu\nu} \,\langle UQU^\dagger Q\rangle + \cdots 
\ee
When combined with the lowest--order $\Delta S=1$ lagrangian, the couplings
\eqn{eq:l4} give rise to physical contributions to the various 
Kaon decays we are going to consider here.
 
Another source of $O(p^4)$ contributions comes from direct 
$\Delta S=1$ terms.
Although the complete list of possible chiral structures 
\cite{KMW:90,EC:90,EF:91,EKW:93} is rather long,
only a few terms are relevant \cite{EPR:87a,EPR:87b,EPR:88}
for the kind of processes we are going to discuss (radiative K decays
with at most one pion in the final state):
\beqn\label{eq:lweak}
\cL_4^{\Delta S=1,\mbox{\rms em}} &\!\!\doteq &\!\!
-{G_F \over \sqrt{2}}  V_{ud}^{\phantom{*}} V_{us}^* \, g_8 
\,\bigg\{
-{i e \over f^2} F^{\mu\nu} \,\left[
   w_1 \,\langle Q \lambda L_\mu L_\nu\rangle  
 + w_2 \,\langle Q L_\mu \lambda L_\nu\rangle \right]
\no\\ &&\qquad\qquad\qquad\;\;\,
\mbox{} + e^2 f^2 w_4\, F^{\mu\nu} F_{\mu\nu}
 \,\langle\lambda QU^\dagger QU\rangle 
 + \mbox{\rm h.c.} \bigg\} .
\eeqn
%

%
%
\newsec{$K\to\pi\nu\bar\nu$}

The decay $K^+\to\pi^+\nu\bar\nu$ is a well--known example of an allowed
process where long--distance effects play a negligible role
\cite{RS:89,HL:89,LW:94,FA:96,GHL:96}. Thus,
this mode provides a good test of the radiative structure of the
Standard Model.
The decay process is dominated by short--distance loops ($Z$ penguin,
$W$ box) involving the heavy top quark, but receives also sizeable
contributions from internal charm--quark exchanges.
The resulting decay amplitude,
\bel{eq:pnn} \mbox{}\!\!  
T(K\to\pi\nu\bar\nu)\,\sim\, \sum_{i=c,t}
F(V_{id}^{\phantom{*}} V_{is}^*;x_i)\;
\left(\bar\nu_L\gamma_\mu\nu_L\right)\;
\langle\pi |\bar s_L\gamma^\mu d_L|K\rangle
\, ,
\qquad\qquad x_i\equiv m_i^2/M_W^2 , 
\ee
involves the hadronic matrix element of the $\Delta S=1$ vector current,
which (assuming isospin symmetry) can be obtained from $K_{l3}$ decays.
In the ChPT framework, the needed hadronic matrix element is known
at $O(p^4)$; this allows to make a reliable estimate of the relevant
isospin--violating corrections \cite{LR:84,MP:96}.

Summing over the three neutrino flavours and expressing the 
quark--mixing factors
through the Wolfenstein parameters \cite{WO:83} $\lambda$, $A$, $\rho$ and
$\eta$, one can write the approximate formula \cite{buras}:
\bel{eq:br_pnn}
\mbox{\rm Br}(K^+\to\pi^+\nu\bar\nu) \,\approx\, 1.93\times 10^{-11}\,
A^4 \, x_t^{1.15}\,\left[ \eta^2 + (\rho_0-\rho)^2\right] \, ;
\qquad\qquad \rho_0\approx 1.4 \, .
\ee
The departure of $\rho_0$ from unity measures the impact of the
charm contribution.

With the presently favoured values for the quark--mixing parameters, 
the branching ratio is predicted to be in the range\cite{buras} 
\bel{eq:pred_br}
\mbox{\rm Br}(K^+\to\pi^+\nu\bar\nu) \,=\,
(9.1\pm 3.2)\times 10^{-11} \, ,
\ee
to be compared with the present experimental upper bound \cite{AD:96}
$\mbox{\rm Br}(K^+\to\pi^+\nu\bar\nu) < 2.4\times 10^{-9}$
(90\% CL).

What is actually measured is the transition $K^+ \to \pi^+ +$ nothing;
therefore,
the experimental search for this process can also be used
to set limits on possible exotic decay modes like $K^+\to\pi^+ X^0$,
where $X^0$ denotes an undetected
light Higgs or  Goldstone boson (axion, familon, majoron, \ldots).
 
The CP--violating decay $K_L\to\pi^0\nu\bar\nu$ has been suggested
\cite{LI:89} as a good candidate to look for  pure
direct CP--violating transitions. 
The contribution coming from indirect
CP violation via $K^0$--$\bar K^0$ mixing is very small \cite{LI:89}:
Br$|_\varepsilon \sim 5 \times 10^{-15}$.
The decay proceeds almost entirely through direct CP violation, and
is completely dominated by short--distance loop diagrams with top
quark exchanges \cite{buras}:
\bel{eq:br_pnn0}
\mbox{\rm Br}(K_L\to\pi^0\nu\bar\nu) \,\approx\, 8.07\times 10^{-11}\,
A^4 \, \eta^2 \, x_t^{1.15} \, .
\ee
The present experimental upper bound \cite{WE:94},
$\mbox{\rm Br}(K_L\to\pi^0\nu\bar\nu) < 5.8\times 10^{-5}$
(90\% CL),
is still far away from the expected range \cite{buras}
\bel{eq:pred_br0}
\mbox{\rm Br}(K_L\to\pi^0\nu\bar\nu) \,=\,
(2.8\pm 1.7)\times 10^{-11} \, .
\ee
Nevertheless, the experimental prospects to reach the required sensitivity
in the near future look rather promising \cite{LI:96,WI:96}.
The clean observation of just a single
unambiguous event would indicate the existence of CP--violating
$\Delta S = 1$ transitions.

%
%
\newsec{$K_S\to\gamma\gamma$}

\begin{figure}[htb]
\vfill
\centerline{
\begin{minipage}[t]{.47\linewidth}\centering
\mbox{\epsfig{file=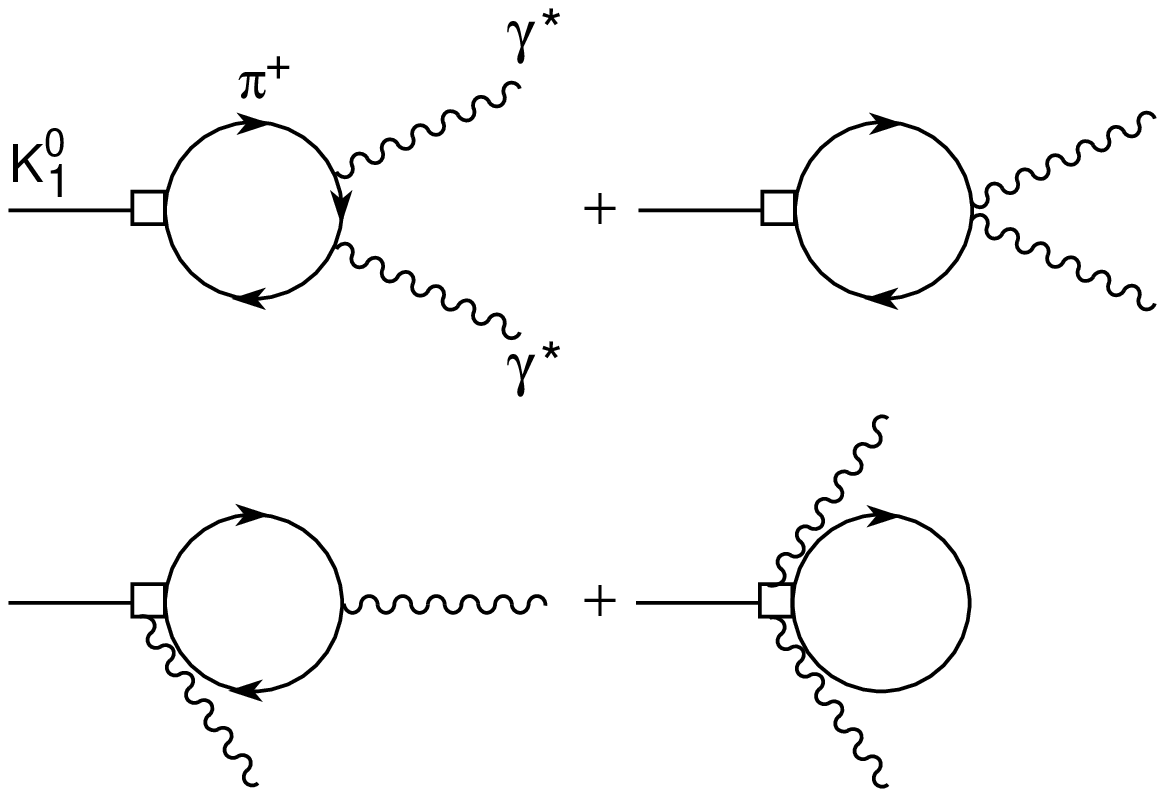,width=8.0cm}}   
\caption{Feynman diagrams for $K_1^0\to\gamma^*\gamma^*$.}
\label{fig:ksgg}
\end{minipage}
\hspace{1.0cm}
\begin{minipage}[t]{.47\linewidth}\centering
\vspace*{-4.5cm}
\mbox{\epsfig{file=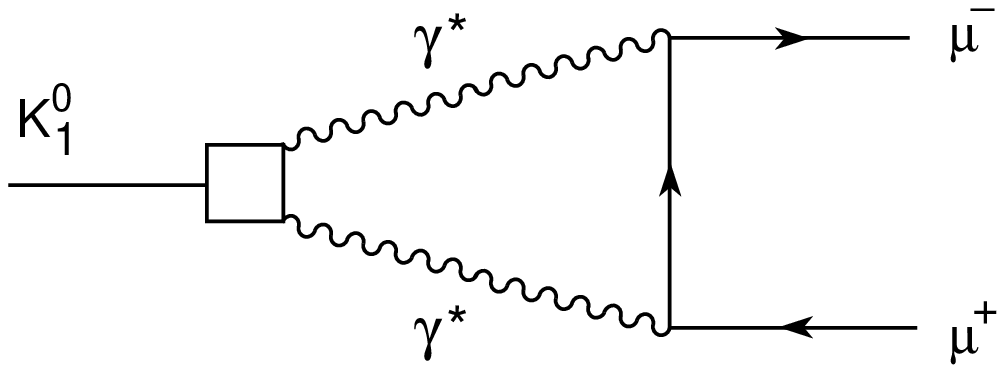,width=7.0cm}}    
\vspace*{0.5cm}
\caption{Feynman diagram for the  decay $K_1^0\to \mu^+ \mu^-$.
The $K_1^0 \gamma^* \gamma^*$ vertex is generated through 
the one-loop
diagrams shown in Fig.~\protect\ref{fig:ksgg}}
\label{fig:ksmm}
\end{minipage}
}
\vfill
\end{figure}

The symmetry constraints do not allow any direct tree--level
$K_1^0\gamma\gamma$ coupling at $O(p^4)$
($K^0_{1,2}$ refer to the CP--even and CP--odd eigenstates, 
respectively).
This decay proceeds then
through a loop of charged pions as shown in
Fig.~\ref{fig:ksgg} (there are
similar diagrams with charged kaons in the loop, but
their sum is proportional to
$M^2_{K^0} - M^2_{K^+}$ and therefore can be neglected).
Since there are no possible counter-terms to renormalize
divergences, the one--loop amplitude is necessarily finite.
Although each of the four diagrams in Fig.~\ref{fig:ksgg}
is quadratically divergent, these divergences cancel in the sum.
The resulting prediction \cite{dAE:86,GO:87},
$\mbox{\rm Br}(K_S\to\gamma\gamma) = 2.0 \times 10^{-6}$,
is in very good agreement
with the experimental measurement \cite{BA:95,BU:87}:
\bel{eq:ksgg}
\mbox{\rm Br}(K_S\to\gamma\gamma) \, = \,
(2.4 \pm 0.9) \times 10^{-6} \, . 
\ee
%
 
%
%
\newsec{$K_{L,S}\to\mu^+\mu^-$}

There are well--known short--distance contributions \cite{buras}
(electroweak penguins and box diagrams)
to the decay $K_L\to\mu^+\mu^-$.
However, this transition is dominated by long--distance
physics. The main contribution proceeds through a two--photon
intermediate state: $K_2^0\to\gamma^*\gamma^*\to\mu^+\mu^-$.
Contrary to $K_1^0\to\gamma\gamma$,
the prediction for the $K_2^0\to\gamma\gamma$ decay is
very uncertain, because the first non-zero contribution
occurs\footnote{
At $O(p^4)$, this decay proceeds through
a tree--level $K_2^0\to\pi^0,\eta$ transition, followed by
$\pi^0,\eta\to\gamma\gamma$ vertices.
Because of the Gell-Mann--Okubo relation,
the sum of the $\pi^0$ and $\eta$ contributions
cancels exactly to lowest order.
The decay amplitude is then very sensitive to $SU(3)$ breaking.}
at $O(p^6)$.
That makes very difficult any attempt to
predict the $K_{L}\to\mu^+\mu^-$ amplitude.
  
The situation is completely different for the $K_S$ decay.
A straightforward chiral analysis \cite{EP:91}
shows that, at lowest order in momenta, the only allowed
tree--level $K^0\mu^+\mu^-$ coupling corresponds to the
CP--odd state $K_2^0$.
Therefore, the $K_1^0\to\mu^+\mu^-$ transition can only be
generated by a finite non-local two--loop contribution.
The explicit calculation \cite{EP:91} gives:
\bel{eq:ksmm_ratios}
{\Gamma(K_S\to\mu^+\mu^-)\over\Gamma(K_S\to\gamma\gamma)}
= 2\times 10^{-6}, \qquad\qquad
{\Gamma(K_S\to e^+ e^-)\over\Gamma(K_S\to\gamma\gamma)}
= 8\times 10^{-9},
\ee
well below the present (90\% CL) experimental upper limits  
\cite{GJ:73,BL:94}:    
Br$(K_S\to\mu^+\mu^-) < 3.2\times 10^{-7}$,
Br$(K_S\to e^+e^-) < 2.8\times 10^{-6}$.
Although, in view of
the smallness of the predicted ratios,
this calculation seems quite academic, it has important
implications for CP--violation studies.

The longitudinal muon polarization $\cP_L$
in the decay $K_L\to\mu^+\mu^-$ is an interesting measure of 
CP violation.
As for every CP--violating observable in the neutral kaon system,
there are in general two different kinds of contributions to $\cP_L$:
indirect CP violation through the small 
$K_1^0$ admixture of the $K_L$
($\varepsilon$ effect), and direct CP violation in the 
$K_2^0\to\mu^+\mu^-$
decay amplitude.
 
   In the Standard Model, the direct CP--violating amplitude is
induced by Higgs exchange with an effective one--loop flavour--changing
$\bar s d H$ coupling \cite{BL:86}.
The present lower bound
on the Higgs mass, $M_H>66$ GeV (95\% CL), implies a
conservative upper limit
$|\cP_{L,\mbox{\rms Direct}}| < 10^{-4}$.
Much larger values, $\cP_L \sim O(10^{-2})$, appear quite naturally
in various extensions of the Standard Model \cite{GN:90,MO:93}.
It is worth emphasizing that $\cP_L$ is especially
sensitive to the presence of light scalars with CP--violating
Yukawa couplings. Thus, $\cP_L$ seems to be a good signature to look
for new physics beyond the Standard Model; for this to be the case,
however, it is very important to have a good quantitative
understanding of the Standard Model prediction to allow us to infer,
from a measurement of $\cP_L$, the existence of a new CP--violation
mechanism.
 
  The chiral calculation of the $K_1^0\to\mu^+\mu^-$ amplitude
allows us to make a reliable estimate
of the contribution to $\cP_L$ due to $K^0$--$\bar K^0$ mixing
\cite{EP:91}:
\bel{eq:p_l}
1.9 \, < \, |\cP_{L,\varepsilon}| \times 10^3 \Biggl( 
{2 \times 10^{-6} \over
\mbox{\rm Br}(K_S\to\gamma\gamma)} \Biggr)^{1/2} \, < 2.5 \, .
\ee
Taking into account
the present experimental errors in $\mbox{\rm Br}(K_S\to\gamma\gamma)$  and
the inherent theoretical uncertainties due to uncalculated
higher--order corrections,
one can conclude that experimental indications for
$|\cP_L|>5\times 10^{-3}$ would constitute clear evidence
for additional
mechanisms of CP violation beyond the Standard Model.

%
%
\newsec{$K\to\pi\gamma\gamma$}

The most general form of the $K\to\pi\gamma\gamma$ amplitude
depends on four independent invariant
amplitudes \cite{EPR:88} A, B, C and D:
%
\beqn\label{eq:a_b_def}
\lefteqn{{\cal A}[K(p_K)\to\pi(p_\pi)\gamma(q_1)\gamma(q_2)]\, =\, 
    \epsilon_\mu(q_1) \,\epsilon_\nu(q_2) \, \Biggl\{ 
    {A(y,z)\over M^2_K}\,
    \Bigl(  q_2^\mu q_1^\nu - q_1\cdot q_2 \, g^{\mu\nu}\Bigr) 
\Biggr. }\no \\
&& \Biggl. \qquad\mbox{}
+ {2 B(y,z)\over M^4_K}\,
\Bigl(p_K\cdot q_1 \, q_2^\mu p_K^\nu
+ p_K\cdot q_2\, q_1^\nu p_K^\mu
- q_1\cdot q_2 \,  p_K^\mu p_K^\nu  -
p_K\cdot q_1\, p_K\cdot q_2 \, g^{\mu\nu}\Bigr)
\Biggr. \no\\
&& \Biggl. \qquad\mbox{}
+ {C(y,z)\over M^2_K}\,\epsilon^{\mu\nu\rho\sigma} q_{1\rho}q_{2\sigma}
\Biggr.\\ && \Biggl.\qquad\mbox{}
+ {D(y,z)\over M^4_K}\,\left[
\epsilon^{\mu\nu\rho\sigma}\left(
p_K\cdot q_2\, q_{1\rho} + p_K\cdot q_1\, q_{2\rho}\right) p_{K\sigma}
+ \left( p_K^\mu \epsilon^{\nu\alpha\beta\gamma} +
  p_K^\nu \epsilon^{\mu\alpha\beta\gamma}\right)
p_{K\alpha}q_{1\beta}q_{2\gamma}\right]
 \Biggr\} , \no
\eeqn
where $y\equiv|p_K\cdot(q_1-q_2)|/M_K^2$ and $z=(q_1+q_2)^2/M_K^2$.
In the limit where CP is conserved, the amplitudes A  and B contribute 
to $K_2\to\pi^0\gamma\gamma$ whereas $K_1\to\pi^0\gamma\gamma$
involves the other two amplitudes C and D. All four amplitudes
contribute to $K^+\to\pi^+\gamma\gamma$.
Only $A(y,z)$ and $C(y,z)$ are non-vanishing to 
lowest non-trivial order, $O(p^4)$, in ChPT.

\begin{figure}[thb]
\centerline{
\begin{minipage}{.47\linewidth}\centering
\mbox{\epsfig{file=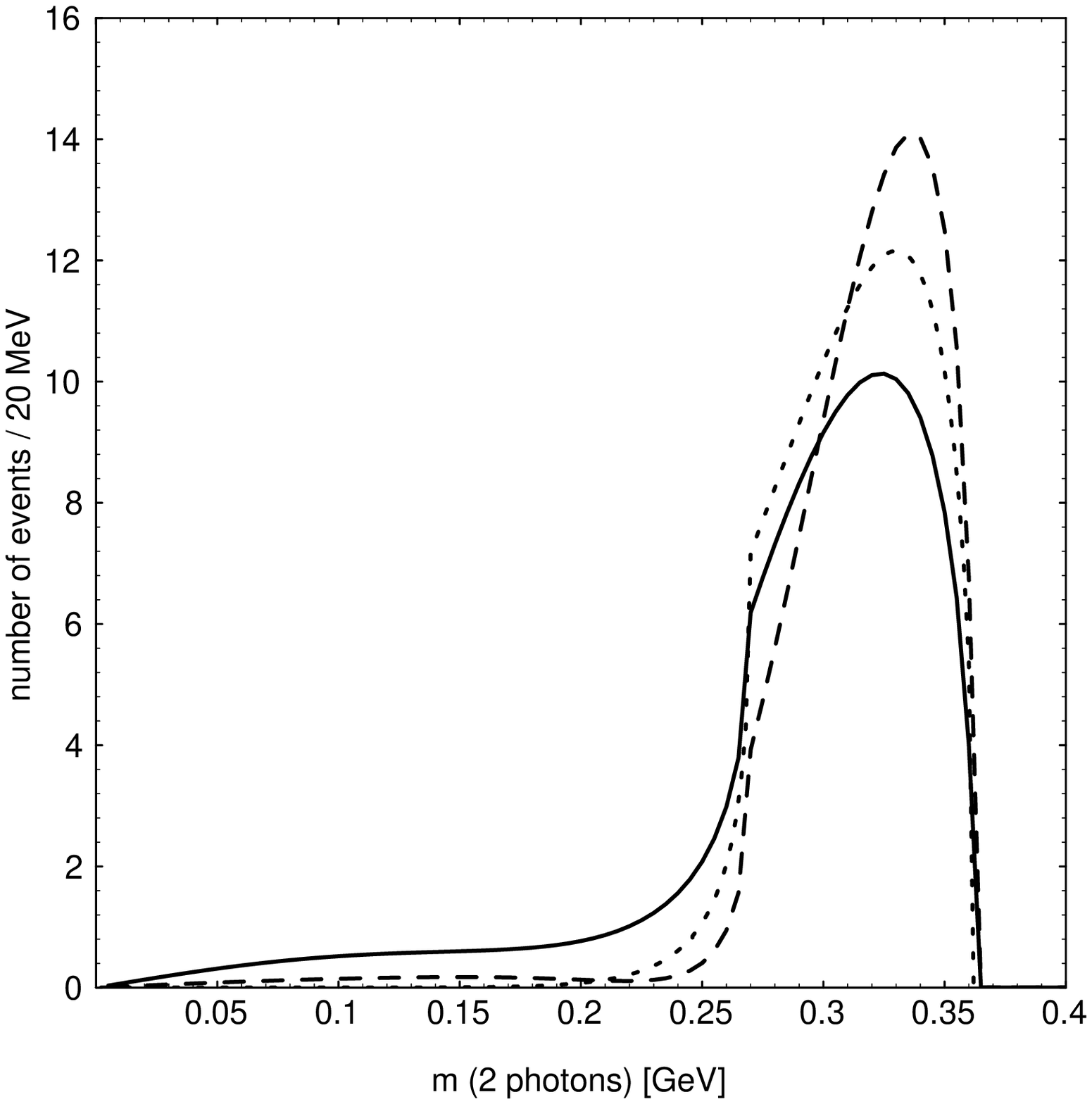,height=7.98cm,width=6.9cm}}  
\caption{$2\gamma$-invariant-mass distribution for
$K_L\to\pi^0\gamma\gamma$:
$\Or (p^4)$ (dotted curve),
$\Or (p^6)$ with $a_V=0$ (dashed curve),
$\Or (p^6)$ with $a_V=-0.9$ (full curve).
The spectrum is normalized to the 50 unambiguous
events of NA31 \protect\cite{BA:92} (without acceptance corrections).}
\label{fig:spectrum}
\end{minipage}
\hspace{0.8cm}
\begin{minipage}{.47\linewidth}\centering
\hbox{}\vspace*{0.9cm}\hspace*{-0.2cm}
\mbox{\epsfig{file=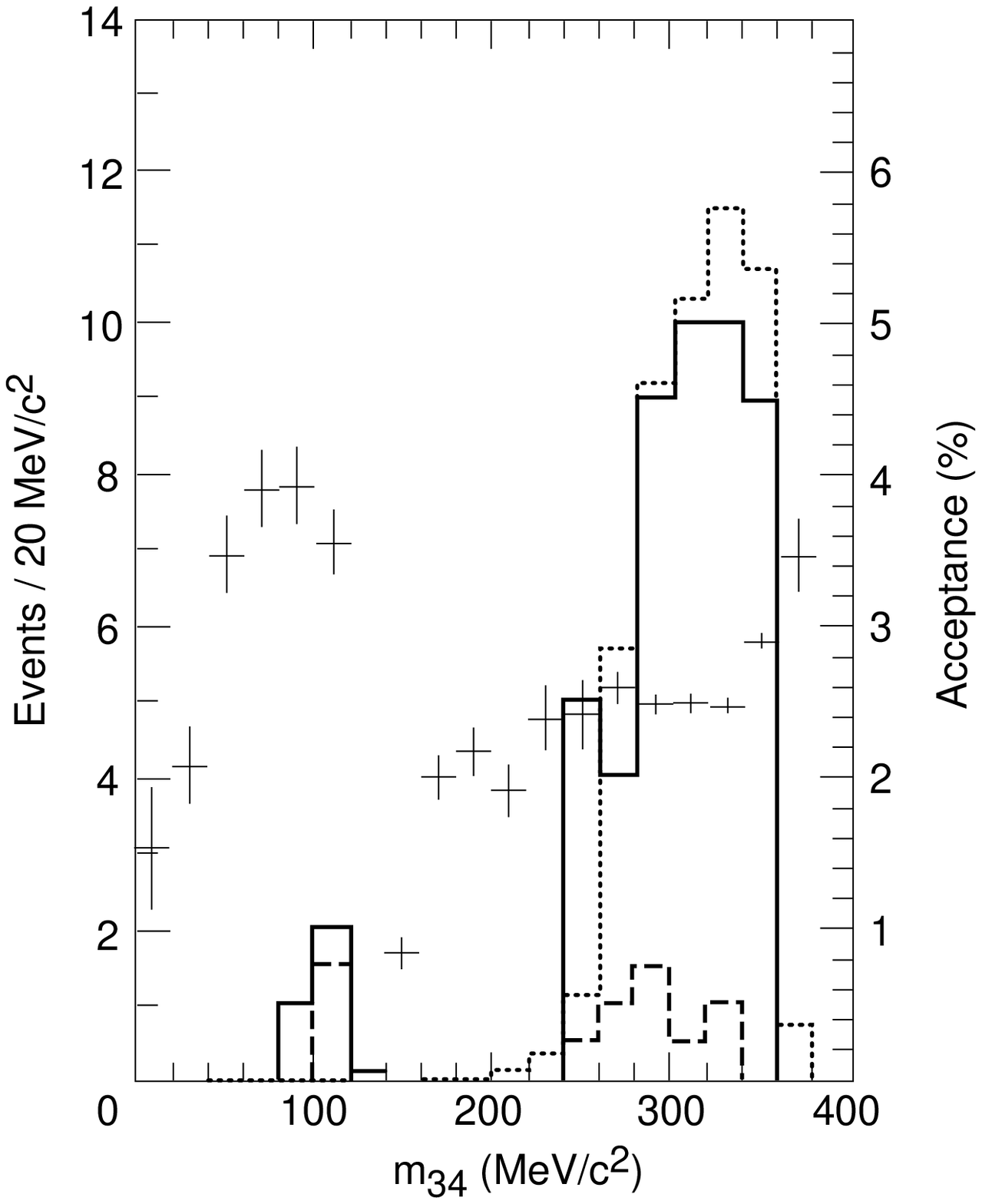,height=6.94cm,width=8.1cm}}
\caption{Measured
\protect\cite{BA:92}
$2\gamma$-invariant-mass distribution for
$K_L\to\pi^0\gamma\gamma$ (solid line).
The dashed line shows the estimated background.
The experimental acceptance is given by the crosses.
The dotted line simulates the $\Or (p^4)$ ChPT prediction.}
\label{fig:spectrum_NA31}
\end{minipage}
}
\vfill
\end{figure}
 
Again, the symmetry constraints do not allow any
tree--level contribution to $K_2\to\pi^0\gamma\gamma$
from $O(p^4)$ terms in the lagrangian.
The $A(y,z)$ amplitude is therefore determined by a
finite loop calculation \cite{EPR:87b}.
The relevant Feynman diagrams are analogous to the ones in
Fig.~\ref{fig:ksgg}, but with an additional $\pi^0$ line
emerging from the weak vertex;
charged kaon loops also give a small contribution in this case.
Due to the large absorptive $\pi^+\pi^-$ contribution,
the spectrum in the invariant mass of the two photons
is predicted \cite{EPR:87b,CdA:88}
to have a very characteristic behaviour
(dotted line in Fig.~\ref{fig:spectrum}),
peaked at high values of $m_{\gamma\gamma}$.
The agreement with the measured two--photon distribution \cite{BA:92},
shown in Fig.~\ref{fig:spectrum_NA31},
is remarkably good.
However, the $O(p^4)$ prediction for the rate \cite{EPR:87b,CdA:88},
$\mbox{\rm Br}(K_L \rightarrow \pi^0 \gamma \gamma) = 0.67\times 10^{-6}$,
is smaller than the experimental value: \cite{BA:92,PA:91}
\bel{eq:br_klpgg}
\mbox{\rm Br}(K_L \rightarrow \pi^0 \gamma \gamma )
\, = \, (1.70 \pm 0.28) \times 10^{-6} \, . 
\ee

Since the effect of the amplitude $B(y,z)$ first appears at
$O(p^6)$, one should worry about the size of the next--order
corrections. A na\"{\i}ve vector--meson--dominance
(VMD) estimate \cite{SE:88,MI:89,FR:89,HS:93}
through the decay chain
$K_L\to\pi^0,\eta,\eta'\to V \gamma\to\pi^0\gamma\gamma$
results \cite{EPR:90} in a sizeable contribution to $B(y,z)$,
\beqn\label{eq:vmd_contribution}
\lefteqn{
A(y,z)\big|_{\mbox{\rms VMD}} \, = \, \tilde a_V 
\left( 3 - z + {M^2_\pi\over M_K^2}\right) ,
\qquad\qquad
B(y,z)\big|_{\mbox{\rms VMD}} \, =\,   -2\tilde a_V \, ,
} \no\\ && \mbox{}\qquad\qquad\qquad
\tilde a_V \,\equiv\, 
-{G_F\over\sqrt{2}} V_{ud}^{\hphantom{*}} V_{us}^*\, g_8 \, 
{M_K^2 \alpha\over\pi} \, a_V \, , \qquad\qquad\qquad\qquad\mbox{}
\eeqn
with $a_V \approx 0.32$.
However, this type of calculation predicts a photon
spectrum peaked at low values of $m_{\gamma\gamma}$,
in strong disagreement with experiment.
As first emphasized in ref.~\citen{EPR:90},
there are also so--called direct weak contributions
associated with $V$ exchange, which cannot be written as a strong
VMD amplitude with an external weak transition.
Model--dependent estimates of this direct contribution \cite{EPR:90}
suggest a strong cancellation with the
na\"{\i}ve vector--meson--exchange effect;
but the final result is unfortunately quite uncertain.
 
A detailed calculation of the most important $O(p^6)$ 
corrections has been performed in ref.~\citen{CEP:93}.
In addition to the VMD contribution, the unitarity corrections
associated with the two--pion intermediate state
(i.e. $K_L\to\pi^0\pi^+\pi^-\to\pi^0\gamma\gamma$) have been
included \cite{CEP:93,CdAM:93}.
Fig.~\ref{fig:spectrum} shows the resulting photon spectrum
for $a_V=0$ (dashed curve) and $a_V=-0.9$ (full curve).
The corresponding branching ratio is:
\bel{eq:br_pred_p6}
\mbox{\rm Br}(K_L\to\pi^0\gamma\gamma) \, = \, \cases{
0.67\times 10^{-6} , & \quad $O(p^4)$, \cr
0.83\times 10^{-6} , & \quad $O(p^6), \, a_V=0\, $, \cr
1.60\times 10^{-6} , & \quad $O(p^6), \, a_V=-0.9\, $. }
\ee
The unitarity corrections by themselves raise the rate only
moderately. Moreover, they produce an even more pronounced
peaking of the spectrum at large $m_{\gamma\gamma}$, which
tends to ruin the success of the $O(p^4)$ prediction.
The addition of the $V$ exchange contribution restores again
the agreement.
Both the experimental rate and the spectrum
can be simultaneously reproduced with  $a_V = -0.9$.
A more complete unitarization of the $\pi$--$\pi$ intermediate
states \cite{KH:94}, 
including the experimental $\gamma\gamma\to\pi^0\pi^0$
amplitude, increases the $K_L\to\pi^0\gamma\gamma$ decay width 
some 10\%, leading to a slightly smaller value of $|a_V|$.

For the charged decay $K^+\to\pi^+\gamma\gamma$, the sum of all
1--loop diagrams gives also a finite $O(p^4)$ amplitude
$A(y,z)$. However, chiral symmetry allows in addition for a direct
tree--level contribution proportional to the 
renormalization--scale--invariant constant \cite{EPR:88}
\bel{eq:c_deff}
\hat c = 32 \pi^2 \left[ 4 \left( L_9 + L_{10}\right)
-{1\over 3} \left( w_1 + 2 w_2 + 2 w_4 \right)\right] .
\ee
There is also a contribution to $C(y,z)$, generated by
the chiral anomaly \cite{EPR:88}.
Since $\hat c$ is unknown, ChPT alone cannot predict 
$\Gamma(K^+\to\pi^+\gamma\gamma)$; nevertheless, it gives, up to
a twofold ambiguity, a precise correlation between the rate and the
spectrum. Moreover, one can derive the lower bound \cite{EPR:88}
Br$(K^+\to\pi^+\gamma\gamma)\geq 4\times 10^{-7}$.

From na\"{\i}ve power--counting arguments one expects $\hat c\sim O(1)$,
although $\hat c = 0$ has been obtained in some models \cite{EPR:90}.
The shape of the $z$ distribution is very sensitive to $\hat c$
and, for reasonable values of this parameter, is predicted \cite{EPR:88}
again
to peak at large $z$ due to the rising absorptive part of the
$\pi\pi$ intermediate state.
The preliminary results of the BNL-E787 experiment
\cite{SH:96,NA:96} show indeed a clear enhancement of events
at large $z$, in nice agreement with
the theoretical expectations.

A recent analysis of the main $O(p^6)$ corrections \cite{dAP:96},
analogous to the one previously performed for the $K_L$ decay mode
\cite{CEP:93,CdAM:93},
suggests that the unitarity corrections generate again a sizeable
($\sim 30$--40\%) increase of the decay width.

%
%
\newsec{$K\to\pi l^+ l^-$}

In contrast to the previous processes,
the  $O(p^4)$  calculation of $K^+\to\pi^+ l^+ l^-$
and $K_S\to\pi^0 l^+ l^-$ involves a divergent loop,
which is renormalized by the $O(p^4)$ lagrangian.
The decay amplitudes can then be written \cite{EPR:87a}
as the sum of a calculable loop
contribution plus an unknown combination of chiral couplings,
\beqn\label{eq:wp_w0}
w_+ &\!\! =  &\!\!  
   -{1\over 3} (4\pi)^2 \left[w_1^r + 2 w_2^r - 12 L_9^r\right]
  -{1\over 3} \log{\left(M_K M_\pi/\mu^2\right)} ,
\no\\
w_S  &\!\! =  &\!\! -{1\over 3} (4\pi)^2 \left[w_1^r - w_2^r\right]
  -{1\over 3} \log{\left(M_K^2/\mu^2\right)} , 
\eeqn
where $w_+$, $w_S$
refer to the decay of the $K^+$ and $K_S$ respectively.
These constants are expected to be of $O(1)$ by
na\"{\i}ve power--counting arguments.
The logarithms have been included to compensate the
renormalization--scale dependence of the chiral couplings,
so that $w_+$, $w_S$ are observable quantities.
If the final amplitudes are required to transform as
octets, then $w_2 = 4 L_9$, implying
$w_S = w_+ + {1\over 3}\log{\left(M_\pi/M_K\right)}$.
It should be emphasized
that this relation goes beyond the usual requirement
of chiral invariance.
 
The measured $K^+\to\pi^+ e^+ e^-$ decay rate determines \cite{EPR:87a}
two possible solutions for $w_+$.
The two--fold ambiguity can be solved, looking to
the shape of the invariant--mass distribution of the final lepton
pair, which is regulated by the same parameter $w_+$.
A fit to the BNL--E777 data \cite{AL:92} gives
\bel{eq:omega}
w_+ = 0.89{\,}^{+0.24}_{-0.14}\, ,
\ee
in agreement with model--dependent
theoretical estimates \cite{EPR:90,BP:93}.
Once $w_+$ has been fixed, one can
predict\cite{EPR:87a}
the rates and Dalitz--plot distributions
of the related modes
$K^+\to\pi^+ \mu^+ \mu^-$,
$K_S\to\pi^0 e^+ e^-$ and $K_S\to\pi^0 \mu^+ \mu^-$.
The preliminary value
Br$(K^+\to\pi^+ \mu^+ \mu^-) = (5.0\pm 0.4\pm 0.6)\times 10^{-8}$,
reported at this workshop by the BNL-787 experiment \cite{SH:96}, 
is in excellent agreement with the theoretical prediction
\cite{EPR:96}
Br$(K^+\to\pi^+ \mu^+ \mu^-) = (6.2{\,}^{+0.8}_{-0.6})\times 10^{-8}$.

%
%
\newsec{$K_L\to\pi^0 e^+ e^-$}

 The rare decay $K_L \rightarrow \pi^0 e^+ e^-$
is an interesting process
in looking for new CP--violating signatures.
If CP were an exact symmetry,
only the CP--even state $K_1^0$ could decay 
via one--photon emission, while
the decay of the CP--odd state $K_2^0$ would proceed through a 
two--photon intermediate state and, therefore,
its decay amplitude would be suppressed
by an additional power of $\alpha$.
When CP violation is taken into account,
however, an $O(\alpha)$ $K_L \rightarrow \pi^0 e^+ e^-$ decay
amplitude is induced, both through the small
$K_1^0$ component of the $K_L$
($\varepsilon$ effect) and through direct CP violation in the
$K_2^0 \rightarrow \pi^0 e^+ e^-$ transition.
The electromagnetic suppression of the CP--conserving amplitude then
makes it plausible that this decay is
dominated by the CP--violating contributions.
 
  The short--distance analysis of the product of 
weak and electromagnetic
currents allows a reliable calculation of the direct CP--violating
$K_2^0 \rightarrow \pi^0 e^+ e^-$ amplitude.
The corresponding branching ratio has been estimated
to be: \cite{buras}
\bel{eq:direct}
\mbox{\rm Br}(K_L \rightarrow \pi^0 e^+ e^-)\Big|_{\mbox{\rms Direct}}
= (4.5\pm 2.6) \times 10^{-12} .
\ee

The indirect CP--violating amplitude induced by the 
$K_1^0$ component of
the $K_L$ is given by the $K_S \rightarrow \pi^0 e^+ e^-$ amplitude
times the CP--mixing parameter $\varepsilon$.
Using the octet relation between $w_+$ and $w_S$,
the determination of the parameter $\omega_+$ in 
\eqn{eq:omega}
implies
\bel{eq:indirect}
\mbox{\rm Br}(K_L \rightarrow \pi^0 e^+ e^-)\Big|_{\mbox{\rms Indirect}}
 \le  1.5 \times 10^{-12}.
\ee
Comparing this  value with \eqn{eq:direct},
we see that the direct
CP--violating contribution is expected to be bigger than the
indirect one. This is very different from the situation in
$K \rightarrow \pi \pi$, where the contribution due to mixing
completely dominates.
 
Using the computed  $K_L\to\pi^0\gamma\gamma$ amplitude,
one can estimate the CP--conserving two--photon exchange contribution
to $K_L\to\pi^0e^+e^-$,
by taking the absorptive part due to the two--photon 
discontinuity as an
educated guess of the actual size of the complete amplitude.
At $O(p^4)$, the $K_L\to\pi^0e^+e^-$ decay
amplitude is
strongly suppressed (it is proportional to $m_e$), owing to the
helicity structure of the $A(y,z)$ term \cite{DHV:87,EPR:88}:
\bel{eq:klpee_a}
\mbox{\rm Br}(K_L \rightarrow \pi^0 \gamma ^* \gamma ^* \rightarrow \pi^0
     e^+ e^-)\Big|_{O(p^4)} \,\sim\, 5 \times 10^{-15} .
\ee
This helicity suppression is, however, no longer true at 
the next order in the chiral expansion. 
The $O(p^6)$ estimate \cite{CEP:93} of the amplitude
$B(y,z)$ gives rise to
\bel{eq:klpee_p6}
\mbox{\rm Br}(K_L \rightarrow \pi^0 \gamma^* \gamma^* \rightarrow 
\pi^0 e^+ e^-)
 \Big|_{O(p^6 )} \,\sim\,
\cases{
0.3 \times 10^{-12}, & \quad $a_V=0\, $, \cr
1.8  \times 10^{-12}, & \quad $a_V=-0.9\, $. }
\ee

Thus, the decay width seems to be dominated by the CP--violating
amplitude, but the CP--conserving contribution could also be
important. Notice that if both amplitudes were comparable
there would be a sizeable CP--violating energy asymmetry between the 
$e^-$ and the $e^+$ distributions \cite{SE:88,HS:93,DG:95}.

   The present experimental upper bound \cite{HA:93},
\bel{eq:klpee_exp}
\mbox{\rm Br}(K_L \rightarrow \pi^0 e^+ e^-)\Big|_{\mbox{\rms Exp}}
 < 4.3 \times 10^{-9} \qquad (90\% \mbox{\rm CL}) ,
\ee
is still far away from the expected Standard Model signal,
but the prospects
for getting the needed sensitivity of around $10^{-12}$ in
the next few years are rather encouraging \cite{LI:96,WI:96}.

\goodbreak

To be able to interpret a future experimental measurement of
the decay rate as a (direct) CP--violating signature,
it is first necessary, however,
to pin down more precisely the actual
size of the three different components of the decay amplitude.
Some possible improvements are:
\bi 
\item 
The size of the indirect CP--violating amplitude in eq.~\eqn{eq:indirect}
uses the octet relation between $w_+$ and $w_S$. 
Although consistent with this assumption,
the explicit calculations of those chiral couplings\cite{BP:93}
do not exclude sizeable deviations which could imply a larger contribution
to the decay amplitude.
A more reliable estimate is then required
\cite{PdR:96}.
\item
A measurement of Br$(K_S\to\pi^0 e^+ e^-)$ would directly determine the
size of the indirect CP--violating amplitude.
To bound this contribution below $10^{-12}$, one needs an experimental
upper bound on the $K_S$ branching ration below $3\times 10^{-10}$,
to be compared with the present value \cite{E621}
Br$(K_S\to\pi^0 e^+ e^-)< 3.9 \times 10^{-7}$ (90\% CL). 
\item 
A careful fit to the $K_L \to\pi^0\gamma\gamma$
data, taking the experimental acceptance into account,
would allow to extract the actual value of $a_V$, and fix the
absorptive contribution to the CP--conserving amplitude.
A better understanding of the dispersive piece 
\cite{SE:88,HS:93,DG:95,CH:67}
is also needed.
\ei
%

%
%
\newsec{The Chiral Anomaly in Non-Leptonic $K$ Decays}
 
The chiral anomaly also  appears in the non-leptonic
weak interactions.
A systematic study of all non-leptonic $K$ decays where
the anomaly contributes at leading order, $O(p^4)$,
has been performed in refs.~\citen{ENP:92} and \citen{ENP:94}.
Only radiative $K$ decays are sensitive to the
anomaly in the non-leptonic sector.
 
The manifestations of the anomaly can be grouped in
two different classes of anomalous amplitudes:
reducible and direct contributions.
The reducible amplitudes arise from the contraction of meson
lines between a weak non-leptonic $\Delta S=1$ vertex and the
Wess--Zumino--Witten  functional \cite{WZ:71,WI:83}.
In the octet limit, all reducible anomalous amplitudes of
$O(p^4)$ can be predicted in terms of the coupling $g_8$.
The direct anomalous contributions are generated through
the contraction of the $W$ boson field between
a strong Green function on one side and the
Wess--Zumino--Witten functional on the other.
Their computation is not straightforward, because of the
presence of strongly interacting fields on both
sides of the $W$.
Nevertheless, due to the non-renormalization theorem
of the chiral anomaly \cite{AB:69},
the bosonized form of the direct anomalous amplitudes
can be fully predicted \cite{BEP:92}.
In spite of its anomalous origin, this contribution
is chiral invariant. The anomaly turns out
to contribute to all possible octet terms of
$\cL_4^{\Delta S=1}$ proportional to the
$\varepsilon_{\mu\nu\alpha\beta}$ tensor.
Unfortunately, the coefficients of these terms
get also non-factorizable contributions
of non-anomalous origin, which cannot be computed
in a model--independent way. Therefore, the final
predictions can only be parametrized in terms of four
dimensionless chiral couplings, which are expected
to be positive and of order one.
 
The most frequent {\it anomalous} decays
$K^+\to\pi^+\pi^0\gamma$ and
$K_L\to\pi^+\pi^-\gamma$ share the remarkable feature that the
normally dominant bremsstrahlung amplitude is strongly suppressed,
making the experimental verification of the anomalous amplitude
substantially easier.
This suppression has different origins: $K^+\to\pi^+\pi^0$ proceeds
through the small 27-plet part of the non-leptonic weak interactions,
whereas $K_L\to\pi^+\pi^-$ is CP violating.
The remaining non-leptonic $K$ decays with direct anomalous contributions
are either suppressed by phase space
[$K^+\to\pi^+\pi^0\pi^0\gamma(\gamma)$, 
$K^+\to\pi^+\pi^+\pi^-\gamma(\gamma)$,
$K_L\to\pi^+\pi^-\pi^0\gamma$, $K_S\to\pi^+\pi^-\pi^0\gamma(\gamma)$]
or by the presence of an extra photon in the final state
[$K^+\to\pi^+\pi^0\gamma\gamma$, $K_L\to\pi^+\pi^-\gamma\gamma$].

%
%

\newsec{Summary}

Rare K decays are an important testing ground of the
electroweak flavour theory. With the improved experimental sensitivity 
expected in the near future, they can provide new signals
of CP--violation phenomena and, perhaps, a window into physics
beyond the Standard Model.

The theoretical analysis of these decays is far from trivial due to
the very low mass of the hadrons involved. The delicate interplay
between the flavour--changing dynamics and the confining QCD interaction
makes very difficult to perform precise dynamical predictions.
Fortunately, the Goldstone nature of the pseudoscalar mesons implies
strong constraints on their low--energy interactions, which can be
analyzed with effective lagrangian methods. 
The ChPT framework incorporates all the constraints implied by the
chiral symmetry of the underlying lagrangian at the quark level,
allowing for a clear distinction between genuine aspects of the Standard
Model and additional assumptions of variable credibility usually related 
to the problem of long--distance dynamics.
The low--energy amplitudes are calculable in
ChPT, except for some coupling constants which are not restricted by 
chiral symmetry. These constants reflect our lack of understanding
of the QCD confinement mechanism and must be determined experimentally
for the time being. Further progress in QCD can only improve our
knowledge of these chiral constants, but it cannot modify the low--energy
structure of the amplitudes.

It is important to emphasize that the experimental verification of the
chiral predictions does not provide a test of the detailed dynamics of 
the Standard Model; only the implications of the underlying symmetries
are being proved. The dynamical information is encoded in the chiral 
couplings. Thus, one needs to derive those chiral constants from the
Standard Model itself, to actually test the non-trivial 
low--energy dynamics.
Although this is a very difficult problem,
the recent attempts done in this direction look quite promising.

\begin{Thebibliography}{99}

\bibitem{LI:96} L. Littenberg, these proceedings.

\bibitem{WI:96} B. Winstein, these proceedings.


\bibitem{buras} 
   G. Buchalla, A.J. Buras and M.E. Lautenbacher, {\it Weak Decays 
     Beyond Leading Logarithms}, \RMP\ (1996)
     [hep-ph/9512380]; \\
   A.J. Buras, these proceedings.

\refjl{WE:79}{S. Weinberg}{Physica}{96A}{1979}{327}

\refjl{GL:85}{J. Gasser and H. Leutwyler}{\NP}{B250}{1985}{465, 517, 539}

\refjl{EC:95}{G. Ecker}{\PPNP}{35}{1995}{1}

\refjl{PI:95}{A. Pich}{\RPP}{58}{1995}{563}

\refbk{dR:95}{E. de Rafael}{Chiral Lagrangians and Kaon CP--Violation}{
  in CP Violation and the Limits of the Standard Model, 
  Proc. TASI'94, ed. J.F.~Donoghue (World Scientific, Singapore, 1995)}

\refbk{EPR:96}{G. Ecker, A. Pich and E. de Rafael}{Rare Kaon Decays in
  Chiral Perturbation Theory}{to appear}

\refjl{PGR:86}{A. Pich, B. Guberina and E. de Rafael}{\NP}{B277}{1986}{197}

\refjl{KMW:91}{J. Kambor, J. Missimer and D. Wyler}{\PL}{B261}{1991}{496}

\refjl{KDHMW:92}{J. Kambor \etal}{\PRL}{68}{1992}{1818}

\refjl{EPR:87a}{G. Ecker, A. Pich and E. de Rafael}{\NP}{B291}{1987}{692}

\refjl{EPR:87b}{G. Ecker, A. Pich and E. de Rafael}{\PL}{B189}{1987}{363}

\refjl{EPR:88}{G. Ecker, A.Pich and E. de Rafael}{\NP}{B303}{1988}{665}

\refjl{KMW:90}{J. Kambor, J. Missimer and D. Wyler}{\NP}{B346}{1990}{17}

\refbk{EC:90}{G. Ecker}{Geometrical aspects of the non-leptonic weak
  interactions of mesons}{in Proc. IX Int. Conf. on the Problems of
  Quantum Field Theory, ed. M.K. Volkov (JINR, Dubna, 1990)}

\refjl{EF:91}{G. Esposito--Far\`ese}{\ZP}{C50}{1991}{255}

\refjl{EKW:93}{G. Ecker, J. Kambor, and D. Wyler}{\NP}{B394}{1993}{101}

\refjl{RS:89}{D. Rein and L.M. Sehgal}{\PR}{D39}{1989}{3325}

\refjl{HL:89}{J.S. Hagelin and L.S. Littenberg}{\PPNP}{23}{1989}{1}

\refjl{LW:94}{M. Lu and M.B. Wise}{\PL}{B324}{1994}{461}

\refbk{FA:96}{S. Fajfer}{Long distance contribution to 
  $K^+\to\pi^+\nu\bar\nu$ decay and $O(p^4)$ terms in CHPT}{hep-ph/9602322}

\refjl{GHL:96}{C.Q. Geng, I.J. Hsu and Y.C. Lin}{\PR}{D54}{1996}{877}

\refjl{LR:84}{H. Leutwyler and M. Roos}{\ZP}{C25}{1984}{91}

\refjl{MP:96}{W.J. Marciano and Z. Parsa}{\PR}{D53}{1996}{R1}

\refjl{WO:83}{L. Wolfenstein}{\PRL}{51}{1983}{1945}

\refjl{AD:96}{S. Adler \etal}{\PRL}{76}{1996}{1421}

\refjl{LI:89}{L.S. Littenberg}{\PR}{D39}{1989}{3322}

\refjl{WE:94}{M. Weaver \etal}{\PRL}{72}{1994}{3758}  

\refjl{dAE:86}{G. D'Ambrosio and D. Espriu}{\PL}{B175}{1986}{237}

\refjl{GO:87}{J.L. Goity}{\ZP}{C34}{1987}{341}

\refjl{BA:95}{G.D. Barr \etal}{\PL}{B351}{1995}{579}      

\refjl{BU:87}{H. Burkhardt \etal}{\PL}{B199}{1987}{139}   

\refjl{EP:91}{G. Ecker and A. Pich}{\NP}{B366}{1991}{189}


\refjl{GJ:73}{S. Gjesdal \etal}{\PL}{44B}{1973}{217}

\refjl{BL:94}{A.M. Blick \etal}{\PL}{B334}{1994}{234}

\refjl{BL:86}{F.J. Botella and C.S. Lim}{\PRL}{56}{1986}{1651}

\refjl{GN:90}{C.Q. Geng and J.N. Ng}{\PR}{D42}{1990}{1509}

\refjl{MO:93}{R.N. Mohapatra}{\PPNP}{31}{1993}{39}


\refjl{CdA:88}{L. Cappiello and G. D'Ambrosio}{\NC}{99A}{1988}{155}

\refjl{BA:92}{G.D. Barr \etal}{\PL}{B284}{1992}{440;
                    {\bf B242} (1990) 523}    

\refjl{PA:91}{V. Papadimitriou \etal}{\PR}{D44}{1991}{573} 

\refjl{SE:88}{L.M. Sehgal}{\PR}{D38}{1988}{808;
                                {\bf D41} (1990) 161}

\refjl{MI:89}{T. Morozumi and H. Iwasaki}{\PTP}{82}{1989}{371}

\refjl{FR:89}{J. Flynn and L. Randall}{\PL}{B216}{1989}{221}

\refjl{HS:93}{P. Heiliger and L.M. Sehgal}{\PR}{D47}{1993}{4920}

\refjl{EPR:90}{G. Ecker, A. Pich and E. de Rafael}{\PL}{B237}{1990}{481}

\refjl{CEP:93}{A.G. Cohen, G. Ecker and A. Pich}{\PL}{B304}{1993}{347}

\refjl{CdAM:93}{L. Cappiello, G. D'Ambrosio and M. Miragliuolo}{\PL}{
                B298}{1993}{423}

\refjl{KH:94}{J. Kambor and B.R. Holstein}{\PR}{D49}{1994}{2346}

\bibitem{SH:96} T. Shinkawa, these proceedings.        

\refbk{NA:96}{T. Nakano}{First Observation of the Decay
   $K^+\to\pi^+\gamma\gamma$}{Proc. Moriond 1996}

\refbk{dAP:96}{G. D'Ambrosio and J. Portol\'es}{Unitarity and vector
  meson contributions to $K^+\to\pi^+\gamma\gamma$}{hep-ph/9606213}

\refjl{AL:92}{C. Alliegro \etal}{\PRL}{68}{1992}{278} 

\refjl{BP:93}{C. Bruno and J. Prades}{\ZP}{C57}{1993}{585}

\refjl{DHV:87}{J.F. Donoghue, B.R. Holstein and G. Valencia}{\PR}{D35}{1987}{
                2769}

\refjl{DG:95}{J.F. Donoghue and F. Gabbiani}{\PR}{D51}{1995}{2187}

\refjl{HA:93}{D.A. Harris \etal}{\PRL}{71}{1993}{3918}

\bibitem{PdR:96} A. Pich and E. de Rafael, to appear.

\bibitem{E621} G. Thomson, these proceedings.      

\refjl{CH:67}{T.P. Cheng}{\PR}{162}{1967}{1734}

\refjl{ENP:92}{G. Ecker, H. Neufeld and A. Pich}{\PL}{B278}{1992}{337}

\refjl{ENP:94}{G. Ecker, H. Neufeld and A. Pich}{\NP}{B413}{1994}{321}

\refjl{WZ:71}{J. Wess and B. Zumino}{\PL}{37B}{1971}{95}

\refjl{WI:83}{E. Witten}{\NP}{B223}{1983}{422}

\refjl{AB:69}{S.L. Adler and W.A. Bardeen}{\PR}{182}{1969}{1517}

\refjl{BEP:92}{J. Bijnens, G. Ecker and A. Pich}{\PL}{B286}{1992}{341} 

\end{Thebibliography}
\end{document}